\documentclass[twocolumn,showpacs,preprintnumbers,amsmath,amssymb]{revtex4}
\usepackage{graphicx}
\usepackage{dcolumn}
\usepackage{bm}
\usepackage{bbm}

\begin{document}

\title{Nematic-Isotropic Transition with Quenched Disorder}
\author{L. Petridis and E. M. Terentjev}
\affiliation{ Cavendish Laboratory, University of Cambridge\\
J J Thomson Avenue, Cambridge CB3 0HE, UK}
\date{\today}

\begin{abstract}
Nematic elastomers do not show the discontinuous, first-order, phase
transition that the Landau-De Gennes mean field theory predicts for
a quadrupolar ordering in 3D. We attribute this behavior to the
presence of network crosslinks, which act as sources of quenched
orientational disorder. We show that the addition of weak random
anisotropy results in a singular renormalization of the Landau-De
Gennes expression, adding an energy term proportional to the inverse
quartic power of order parameter $Q$. This reduces the first-order
discontinuity in $Q$. For sufficiently high disorder strength the
jump disappears altogether and the phase transition becomes
continuous, in some ways resembling the supercritical transitions in
external field.\\
\\
to be published on \emph {Phys. Rev. E}
\end{abstract}


\maketitle
\section{Introduction}

Although there is a large volume of literature devoted to the
effects of quenched disorder, there has been relatively little study
on how it influences the behavior of systems whose pure versions
undergo a first-order phase transition. This question was first
addressed by Imry and Wortis \cite{imry79} who showed that
inhomogeneities may cause local variations of the transition
temperature inside the sample. Provided that the cost in interface
energy is not great, bubbles of the ``wrong" phase are formed,
eventually leading to a substantial rounding of the transition. A
theorem due to Aizenman and Wehr \cite{aizenman89} shows that in
less than two dimensions there can be no phase coexistence at the
transition, and therefore no latent heat, in a system with quenched
random impurities. Therefore these systems are expected to always
exhibit a continuous transition.

The influence of quenched impurities coupling to the local energy
density has been extensively studied by Cardy
\cite{cardy96,cardy99}. A mapping to the random field Ising model,
whose renormalization group flows are well known, addresses the
question of what happens in higher than two dimensions where the
Aizenman-Wehr theorem is not applicable. It was found that,
depending on the specific values of parameters such as the laten
heat and the surface tension, the phase transition can either be
first or second order. More conclusive analytic results were
obtained in an Ising model with discrete order parameter
\cite{cardy96}. The pure system exhibits a fluctuation-driven
first order transition, where the mean field theory predicts a
continuous transition but fluctuation effects make it
discontinuous. Quenched randomness eventually drives the
transition to become continuous in two dimensions, in accordance
with the Aizenman-Wehr theorem, but this may or may not happen in
higher dimensions.

The majority of studies in this field are carried out for
spin-glass or analogous systems, e.g. with a frustrated dipolar
ordering \cite{chudnovsky86, feldman00}. However, in such systems
the experimental work is difficult and results are often indirect.
In contrast, the quadrupolar orientational ordering of nematic
liquid crystals offers the easy experimental access to
thermodynamic and structural features of phase transitions. The
classical work on frustrated nematic liquids by Bellini et al.
\cite{bellini95, bellini00} has generated a large interest in
studies of liquid crystals in random environments, such as porous
silica gel. One must appreciate, however, that the characteristic
length scale of such disorder is much greater than the coherence
length of the nematic order parameter and thus the theoretical
concept of a continuous coarse-grained random field \cite{imry75}
is difficult to sustain.

Quenched disorder is intrinsically present in nematic elastomers
as a direct result of their synthesis \cite{fridrikh99,emtbook}.
Sources of disorder are introduced by crosslinking a
liquid-crystalline polymer melt. In the simplest situation, the
crosslinking takes place in the isotropic phase, in which case the
local anisotropy axis of each crosslinking moiety is randomly
oriented. Once the polymer network is formed, the configuration of
the crosslinks remains quenched, that is, it does not change with
time and temperature. Unless special precautions are taken during
network fabrication, the low temperature (ordered) phase of
nematic elastomers is always an equilibrium polydomain director
texture \cite{clarke98,elias99}. This is in marked contrast with a
kinetic ``polydomain'' texture often referred to as Schlieren
texture \cite{degennes95}, which is the consequence of nucleation
and growth mismatch in a system undergoing the first-order
transition \cite{yurke91}. The equilibrium polydomain structure of
nematic elastomers is reversible with changing temperature and is
characterized by the uniform non-zero order parameter, but the
highly non-uniform orientation of the principal axis of nematic
director $\bm{n}$. Correlations between directors decay rapidly
and eventually vanish at distances much larger than $\xi$, the
correlation length or domain size. This is in agreement with the
general result that quenched impurities destroy long-range order,
first shown by Larkin \cite{larkin70} and then generalized by Imry
and Ma \cite{imry75}. There is a full analogy with a corresponding
dipolar system named ``random anisotropy magnets''
\cite{chudnovsky82,cleaver97,fridrikh97}. In fact, all other (e.g.
smectic \cite{olmsted96}) liquid crystal elastomers follow the
same pattern of forming the equilibrium textures with a
characteristic length scale often referred to as the domain size.

This length scale in typical nematic elastomers is of the order of
microns \cite{clarke98,elias99}, therefore light passing through
the sample is multiply scattered on birefringent domains with
randomly oriented optical axis \cite{stark97} (see
\cite{fridrikh99} for a brief review of experimental facts in this
area). As a result such a sample is completely transparent at high
temperatures, but becomes opaque below its nematic-to-isotropic
transition temperature $T_{NI}$. It should be noted that the
polydomain texture is the thermodynamically stable low-temperature
phase. Applying an adequately strong aligning stress
\cite{clarke98} or increasing the temperature above $T_{NI}$
\cite{elias99} destroys the polydomain texture, by aligning the
local axis of each domain, or removing the optical contrast
between them. However, once the stress is removed (or the
temperature lowered), the elastomer returns to its previous state
and the average domain size is found to be reversible during this
stress (or temperature) cycling.

Experimental investigations in nematic elastomers support the
theoretical ideas of a continuous nematic-isotropic transition,
rather than the discontinuous first-order transition expected by
the Landau-De Gennes mean field theory based on the symmetry of
quadrupolar ordering in three dimensions. The fact that polydomain
nematic textures are optically opaque creates a practical problem
when attempting to experimentally determine the local order
parameter $Q$ of the mesogenic units. It is impossible to use
birefringence, dichroism or Xray measurements, the methods that
have made this task simple in aligned liquid-crystal systems.
Nuclear magnetic resonance (NMR) provides perhaps the only
opportunity of order detection, by tracking the bias in
orientational motion of selected chemical bonds and providing a
unique $Q$-signature even in the orientationally averaged case. It
was used to measure $Q(T)$ of both a nematic polymer melt and its
corresponding crosslinks network \cite{disch94,lebar05}. The
addition of crosslinks was shown to make the nematic transition
smooth, as well as to slightly reduce order. An indirect
alternative to NMR comes from applying an external field to align
the domains. Birefringence can then be measured, as long as the
system has passed the critical point of the polydomain-monodomain
transition \cite{fridrikh97}. The order parameter can thus be
found for a series of decreasing applied fields. Although the
zero-stress limit is not accessible, it can be extrapolated from
the comparison of the other curves and, again, a continuous phase
transition is seen \cite{kaufhold91}.

The continuous transition is also found in carefully synthesized
monodomain nematic elastomers, where a second stage of the
crosslinking takes place when the sample is stressed
\cite{kupfer91}. Plots of the macroscopic order parameter $Q(T)$,
obtained through birefringence \cite{finkelmann01}, X-ray
scattering \cite{clarke01} and NMR measurements \cite{lebar05},
show the same behavior. There are two possible explanations for
such a deviation from the basic symmetry-based expectation of a
first-order transition. Inhomogeneities may cause local variations
of the transition temperature inside the sample and lead to a
substantial rounding of the transition, as discussed in
\cite{imry79}. The other explanation considers stresses imprinted
in the system during the second crosslinking stage, which add a
$-fQ$ term to the Landau-De Gennes expansion. Obviously, for
adequately large $f$, the system would become supercritical and
show a continuous transition. The analysis of the NMR spectra
supports the supercritical scenario, although some degree of
inhomogeneity was also found using the second moment of the
spectra \cite{lebar05}. Another study of strain as a function of
temperature over a range of applied tensile stress argues the
opposite point \cite{selinger02}. The blurring of the
isotropic-nematic transition is attributed to the presence of
heterogeneities, although the authors do not consider boundary
effects that are bound to be present as discussed in
\cite{imry79}.

Uchida \cite{uchida00} has studied disordered polydomain nematic
elastomers with emphasis on the role of nonlocal elastic
interactions. He has shown that networks crosslinked in the
isotropic phase lose their long-range orientational order due to
the locally quenched random stresses, which were incorporated into
the affine-deformation model of nematic rubber elasticity.
Simulation work was carried out to investigate the role of random
bonds and random fields that might be present in elastomers
\cite{selinger04}, in both cases finding that the first-order
isotropic-nematic transition to broaden into a smooth crossover.
For random-field disorder, the smooth crossover into an ordered
state is also attributed to the long-range elastic interaction
present in elastomers. A recent coarse-grained model for
liquid-crystalline elastomers has also found that both homogeneous
and inhomogeneous samples undergo a continuous isotropic-nematic
transition \cite{pasini05}.

In this paper we apply the traditional spin-glass techniques to
investigate the characteristics of nematic phase transition in a
system with quenched random anisotropy. The structure of the paper
is as follows. In Section~$II$ we summarize a physical model of
quenched disorder in nematic systems following \cite{fridrikh97},
and introduce the replica Hamiltonian. Section~$III$ applies the
auxiliary fields to incorporate several constraints into this
problem and obtains the effective mean-field free energy of disorder
in the system. In Section~$IV$ we investigate the stability of
replica symmetry and discover the limits where our solutions are
valid. Finally, in Section~$V$ we obtain the final free energy
renormalization in terms of the order parameter $Q$ and disorder
strength, and investigate the characteristics of the nematic phase
transition in various situations. We conclude by discussing our
results and comparing them with experiments.

\section{Model}

\subsection{Sources of Quenched Disorder}

In the case of nematic elastomers crosslinked in the isotropic
phase, the sources of quenched disorder are provided by the
network crosslinks. Almost independently of their specific
chemical structure, the crosslinks contain anisotropic groups that
locally provide easy anisotropy axes $\bm{k}$: it is favorable for
the local director to align along $\bm{k}$ in the vicinity of the
crosslink because the anisotropic molecules in both the
crosslinkers and in the nematic system interact, both sterically
and via the long-range van de Waals attraction. The local
anisotropy axes of the crosslinks, as well as their distribution
inside the sample, are quenched variables since the crosslinks can
neither rotate nor move once the chemical synthesis of the
elastomer is complete. Although this has never been tested
experimentally, there are two independent molecular models that
estimate the energy of orientational confinement that a crosslink
experiences from the surrounding network strands
\cite{verwey97,popov98}

\begin{figure}[tbp]
\begin{center}
\resizebox{0.30\textwidth}{!}{\includegraphics{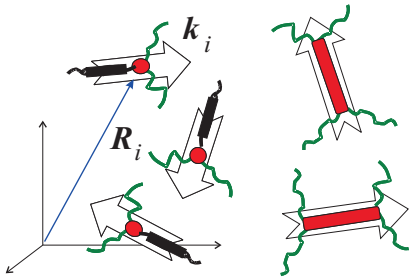}}
\end{center}
\caption{Schematic representation of how crosslinks provide easy
anisotropy axes $\{\mathbf{k} \}$. The nematic director is forced
to be aligned, in the vicinity of the crosslink, with the axes,
which are represent by the arrows. Both the orientation of
$\{\mathbf{k}\}$ as well as the positions of the crosslinks
$\{\mathbf{R}_i\}$ are random. Since the crosslinks are confined
by the network topology, they add quenched disorder to the nematic
system.  } \label{crosslinks}
\end{figure}

We follow earlier work \cite{fridrikh99} in modelling the local
coupling of the nematic order and the random field applied by the
crosslinks. For a crosslink positioned at $\bm{R_i}$, with an
anisotropy axis $\bm{k_i}$, an energy $-\gamma \, \bm{k_i}\cdot
\underline{\underline{\bm{Q}}}\cdot\bm{k_i}$ is raised due to the
interaction with the local nematic order parameter
$Q_{ij}\,=\,Q\left(n_in_j\,-\,\delta_{ij}/3\right)$, where
$\gamma$ is the coupling strength. Employing a coarse-grained
expression for the continuum density of crosslinks $
\rho\left(\bm{r} \right)\,=\,\sum_{\bm{R_i}}\, \delta
(\bf{\bm{r}-\bm{R_i}})$ and substituting the full tensor
expression for $\underline{\underline{\bm{Q}}}$, we get the random
field energy:
\begin{eqnarray}
F_{\rm r.f.}=\, - \, \int{d^3\bm{r}\, \gamma\,Q \,\rho (\bm{r})
\left( \bm{k \cdot n}\right)^2},
 \label{rfenergy}
\end{eqnarray}
where the irrelevant constant $\gamma \int \rho(\bm{r})$ has been
dropped. Although the random field energy is proposed specifically
for a nematic elastomer, it is a general expression which is valid
for all random-anisotropy systems with underlying quadrupolar
symmetry.

To obtain the full Hamiltonian describing the nematic ordering,
the gradient elasticity penalizing the fluctuations of the
director field must be also taken into account. In continuum
elasticity the Hamiltonian takes the form:
\begin{eqnarray}
H[\rho,\bm{k}] \,=\,\int {d^3 \bm{r} \left[ \frac{K}{2}\,\left(
\nabla \bm{n}\right)^2\,-\,\gamma\,Q\,\rho \left(\bm{k} \cdot \bm{n}
\right)^2\right]},
 \label{hamiltonian}
\end{eqnarray}
where $K$ is the Frank elastic constant in the one-constant
approximation. A simple dimensional argument gives $K \sim
k_BT/a$, where $a$ is the nematic coherence length, below which
the meanings of the director $n$ and order parameter $Q$ are
ill-defined \cite{degennes95}. Both microscopic and
phenomenological theories of nematic-isotropic transition give the
elastic constant $K$ to scale as $Q^2$ for $Q \ll 1$. Combining
these two estimates, we take that for small $Q$ the elastic
constant is approximately given by $K \simeq k_BTQ^2/a$. It is
important to clarify that we are examining a homogeneous sample
and as a result the magnitude of $Q$ and the $T_{NI}$ are uniform
across the sample. There is a rich literature on the role
inhomogeneities play in general first-order systems \cite{imry79}
and more specifically in nematic elastomers
\cite{selinger02,uchida00,selinger04,xing03}. The important
difference in our assumptions is that, although the director
correlations in the polydomain nematic are short-range
(equilibrium spin-glass texture), the underlying nematic order
parameter $Q$ is homogeneous across the system. This assumption is
based on the fact that the spin-glass like nematic textures are in
fact homogeneous, in the sense that every element of the sample is
equivalent to others: no `domain walls' (unlike, for instance,
during the stress-induced poly-monodomain transition
\cite{fridrikh99} when the domain walls localize).

\subsection{The Replica Method}
There are three established methods of dealing with quenched
random fields in the replica method framework: one based on the
functional renormalization group analysis, another using the
Gaussian variational method and the third using auxiliary fields.
This paper employs the latter.

We are interested in results that do not depend on the specific
distributions of $\rho(\bm{r})$ and $\bm{k}$ because we cannot
control these distributions experimentally. In other words we are
looking for the free energy averaged over the random distributions
of the quenched variables $\rho(\bm{r})$ and $\bm{k}$.
Crosslinking the sample above $T_{NI}$ makes the easy anisotropy
axes point at random directions, with an isotropic probability of
orientation $P(\bm{k})\,=\,\frac{1}{4\pi}$. Furthermore, the
crosslinks are dispersed randomly in the sample with density
$\rho$. The probability that a particular distribution
$\rho(\bm{r})$ occurs is Gaussian:
\begin{eqnarray}
P[\rho(\bm{r})]\,\sim\, \exp\left[- \int{d^3r \frac{\left[
\rho(\bm{r})- \rho_0 \right]^2}{2\rho_0}}\right], \label{rho}
\end{eqnarray}
where $\rho_0$ is the mean density of crosslinks.

It is not possible to directly average the logarithm of the
partition function $Z$ and obtain the exact free energy. So an
alternative definition of a logarithm (the limit: $\log
Z\,=\,\partial_m Z^m|_{m \rightarrow 0}$) is used, allowing to
perform the simpler average of the product $Z^m$. This way of
dealing with quenched disorder is called the ``replica trick", first
introduced in the context of spin glasses by Edwards and Anderson
\cite{edwards75}. The expression for the free energy arising from
disorder then reads:
\begin{eqnarray}
F_d\,&=&\,-k_BT\,\langle \log Z \rangle _{\rho, \bf k}\, =\,\left. {
- k_B T\frac{\partial }{\partial m}\left\langle {Z^m } \right\rangle
} \right|_{m \rightarrow 0}  \label{replicatrick} \\
 &=&\,-k_BT \left. \frac{\partial}{\partial
m}\right|_{m \rightarrow 0} \prod_{a=1}^m
\int\emph{D}\bm{n}_a\,\exp \left[-\beta H_{\rm rep} \right]
\nonumber
\end{eqnarray}
where we now have $m$ identical``replicas" of the system, labelled
by the index $a$. The aim of this work is to obtain $F_d$ as a
function of the order parameter $Q$ and add it to the Landau-De
Gennes free energy to see how it influences the phase transition.

A rough sketch of the averaging over disorder is given below. The
density average over $P(\rho)$ yields a random field term
 $$
 \sim
\exp\left[\sum_{a,b}(\beta\gamma Q)^2\rho_0
(\bm{k}_a\cdot\bm{n}_a)^2(\bm{k}_b\cdot\bm{n}_b)^2 \right],
$$
where (here and throughout this paper) $\beta=1/k_BT$. The
reader's attention is drawn to the appearance of a second replica
index $b$ due to the square of the $\sum_a$. Since the
distribution of the orientations of the easy axes $P(\bm{k})$ is
assumed fully isotropic, symmetry arguments show that the average
$\langle k_i k_j k_lk_m\rangle_{P(k)}$ is proportional to
$(\delta_{ij}\delta_{lm}+\delta_{il}\delta_{jm}+
\delta_{im}\delta_{jl})$. The constant of proportionality is equal
to $1/(2d+d^2)$, where $d$ is the dimensionality of the unit
vector $\bm{k}$. Therefore
\begin{eqnarray}
&&\langle e^{k^ik^jk^lk^m\,n^in^jn^ln^m}\rangle_{P(k)}=
\nonumber\\
&&=1+\langle {k^ik^jk^lk^m}\rangle\,n^in^jn^ln^m+\,\dots
\nonumber \\
&&\approx e^{\langle k^ik^jk^lk^m\rangle \,n^in^jn^ln^m}.
\label{kaverage}
\end{eqnarray}
Since $(n_a\cdot n_b)\leq 1$ higher terms of the Taylor expansion
are smaller than the lowest order term  ($(n_a\cdot n_b)^2$) and
are subsequently dropped in the last line of Eq.~(\ref{kaverage}).
This approximate treatment retains a random field term which is
overall fourth order in $\bm n$, as it was before the $\bm
k$-averaging, and it is most frequently met, and used, in
molecular theories involving rotational diffusion.

The second line of Eq.~(\ref{replicatrick}) is obtained after
averaging over quenched disorder, sketched above, and  provides
the definition of the ``Replica Hamiltonian'', which no longer
depends on the quenched distributions of $\{\bm{k}\}$ and
$\rho(r)$, but instead couples different replicas of the system:
\begin{eqnarray}
H_{\rm
rep}\left[\bm{n}\left(\bm{r}\right)\right]\equiv\sum_{a,b=1}^m
\int d^3 \bm{r} \bigg\{ \frac{K}{2}\,\left( \nabla
\bm{n}_a\right)^2\delta_{ab}+ \nonumber\\-\,\frac{\Gamma}{2}
\left[2\left(\bm{n}_a \cdot \bm{n}_b \right)^2\,+\,(\bm{n}_a \cdot
\bm{n}_a)\,( \bm{n}_b \cdot \bm{n}_b)\right] \bigg\},
\label{replicahamiltonian}
\end{eqnarray}
where subscripts $a$ and $b$ are the replica indexes and $m$ is the
number of replicas that will be set to zero at the end of the
calculation. Parameter $\Gamma$, arising from completing the
Gaussian square between the Eq.~(\ref{rho}) and the random-field
term in the Eq.~(\ref{hamiltonian}), reflects the strength of the
disorder and has a quadratic dependance on the order parameter:
\begin{eqnarray}
\Gamma \,=\,\frac{\gamma^2\rho_0} {15\,k_BT}\,Q^2. \label{Gamma}
\end{eqnarray}
It is noted that all replicas are assumed to have equal disorder
strength and equal magnitude of the local order parameter, i.e.
$\gamma_a=\gamma_b$ and $Q_a=Q_b$ for all $ a$ and $b$. Bearing in
mind that the director is a unit vector we see that the term with
$(\bm{n}_a \cdot \bm{n}_a)(\bm{n}_b \cdot \bm{n}_b)$ in the random
field part of the Replica Hamiltonian just adds an irrelevant
constant to the expression. We drop this term and keep the
relevant contribution $-\Gamma \left(\bm{n}_a \cdot \bm{n}_b
\right)^2$ that describes the coupling between different replicas.

\section{Disorder Free Energy}
\label{section_3}

\subsection{Auxiliary Fields}

Care must be taken to ensure that the director $\bm{n}(\bm{r})$
remains a unit vector. Although this was assumed to be the case, it
has not been implemented explicitly in the
Eq.~(\ref{replicahamiltonian}). One way to achieve this multiplies
the partition function in Eq.~(\ref{replicatrick}) with the
delta-function constraint
\begin{eqnarray}
\delta\left(\bm{n}^2\,-\,1\right)\,=\frac{1}{2\pi}\,\int_{-\infty}^\infty
\,d\phi \, e^{-i\,\phi\left(\bm{n}^2\,-\,1\right)},
\end{eqnarray}
where $\phi$ is an auxiliary field that allows the delta-function
to be written in its exponential form. We proceed with a
mean-field treatment of the auxiliary fields where a constant
value for $\phi$ is assumed, independent of spacial coordinates.
This approximation implies that the constraint $\bm n^2=1$ is
equally enforced across the whole sample. It is reasonable to
expect this, given that the sample is spatially homogeneous. The
same reasoning explains why $\phi$ has no dependance on the
replica indexes: since its value is independent of the position of
the crosslinks in the sample, it cannot have different values for
different replicas. The corresponding quadratic term $\sum_a
i\,\phi \left(\bm{n}_a^2\,-\,1\right)$ has to be added to the
Replica Hamiltonian in Eq.~(\ref{replicahamiltonian}).

To obtain the disorder energy one must evaluate the statistical
sum over all possible trajectories $\bm{n}_a$. The standard way to
evaluate Hamiltonians with quartic interactions is to introduce an
auxiliary field, here a tensor $\lambda_{ab}$, which reduces the
Hamiltonian to bilinear order in $\bm{n}_a$. To employ the method
all the quantities in $\beta H_{\rm rep}[\bm{n}(\bm{r})]$ must be
dimensionless. Since the integral over $r$ has the dimensions of
volume, we move to a discrete summation over all points in space:
\begin{eqnarray}
\int_a^L dx\, \int_a^L dy\ \int_a^L dz \,=\, a^3\,
\sum_{\rm{points}\ \bm{r}} \nonumber.
\end{eqnarray}
The limits of the $r$-integration are $L$, the size of the system,
and the short-distance cutoff $a$ -- the nematic coherence length,
below which the continuum representation is no longer applicable.
The $(a,b)$ replica coupling term then becomes
\begin{eqnarray}
&& \exp\left[\beta \Gamma a^3\sum_r\left(\bm{n}_a \cdot \bm{n}_b
\right)^2\right]  \label{Fd_lambda}  \\
&=& \int d\lambda_{ab}\exp\left\{\sum_r\left[
-\frac{\lambda_{ab}^2}{4\tilde{\Gamma}}+\lambda_{ab}(\bm{n}_a
\cdot \bm{n}_b) \right]\right\}, \nonumber
\end{eqnarray}
which involves the dimensionless constant $\tilde{\Gamma}=\beta
\Gamma a^3$. It is important to clarify the meaning of the
$\lambda_{ab}^2$ term in Eq.~(\ref{Fd_lambda}): it is the square
of the value $\lambda_{ab}$ rather than an element of the product
of two matrices. Furthermore, from now on we shall use a
mean-field approximation, where it is assumed that $\lambda_{ab}$
has no $r$-dependance. This is an important limitation, but we
believe it is reasonable as we are looking for homogeneous
ordering in the system. Summation over $r$ of the $\lambda_{ab}^2$
term yields $N\,\lambda_{ab}^2/ (4\Gamma)$, where $N=V/a^3$ is the
number of ``discrete spacial points" and $V=L^3$ is the system's
volume.

Moving to the corresponding discrete Fourier space, the effective
Replica Hamiltonian includes both auxiliary fields:
\begin{eqnarray}
&&\beta H_{\rm eff}\left[
\bm{n}\left(q\right)\right]\,=\,\sum_{a,b}\bigg\{ -i \phi\,
\delta_{ab}\,+\, N\frac{
\lambda_{ab}^2}{4\tilde\Gamma}+ \nonumber \\
&+&\, \sum_q \left[ \left(\frac{\tilde{K}q^2}{2}+i\phi\right)
\delta_{ab} \,-\,\lambda_{ab}\right] \left(\bm{n}_a \cdot \bm{n}_b
\right)\bigg\}, \label{Heff}
\end{eqnarray}
with the dimensionless elasticity constant $\tilde K= \beta K
a^3$. The discrete sum over $q$ is related to the integral via
$\sum_q=L^3\int \frac{d^3q}{(2\pi)^3}$. As mentioned above, the
conversion from the integral to the discrete sum is essential so
that all the quantities in $H_{\rm eff}$ (such as $\tilde\Gamma$,
$ \phi$, $ \lambda_{ab}$ and $\tilde K q^2$) remain dimensionless
and the logarithm of their sum can be evaluated correctly.

To be able to deduce the disorder energy as a function of the
order parameter, we make explicit the dependence of $\tilde\Gamma$
and $\tilde K$ on the magnitude of the order parameter $Q$:
\begin{eqnarray}
&&\tilde \Gamma=gQ^2\, , \quad {\rm with} \quad
g=\frac{(\gamma\beta)^2\rho_0 a^3}{15}
\nonumber \\
&&\tilde K = \beta\kappa a^3 Q^2\,, \label{GofQ}
\end{eqnarray}
where $\kappa\sim k_BT/a$ is the Frank elastic constant deep
inside the nematic phase. It is worth noting that $\tilde\Gamma$
is always significantly smaller than one. The distance between
crosslinks, $d_c$, which can be deduced from the crosslink density
$\rho_0\simeq d^{-3}_c$, is found around $7-10$nm in nematic
elastomers \cite{fridrikh99}. A typical coherence length for
nematics is $a\sim 5$nm, hence $(a/d_c)^3\simeq 0.4$ or less. The
coupling of disorder to the nematic director is deduced from the
size of the domains in Ref~\cite{fridrikh99} and is found to be
$\gamma\simeq 0.4k_BT$. Therefore this crude estimate gives
$g\simeq 4\times10^{-3}$ for the nematic elastomers studied in the
literature.

\subsection{Replica Symmetry Case}

To be able to advance in the calculation, a matrix form of the
auxiliary field $\lambda_{ab}$ has to be postulated. A reasonable
starting point is to assume that it has the simplest possible
form, where its elements have a constant value independent of $a$
and $b$:
\begin{eqnarray}
\lambda_{ab}\,=\,\lambda
\left(\mathbbm{1}_{ab}\,-\,\delta_{ab}\right), \label{RSlambda}
\end{eqnarray}
where $\mathbbm{1}_{ab}$ is the matrix with all its elements equal
to one and $\delta_{ab}$ is the identity matrix. This scheme is
frequently encountered in the literature \cite{dotsenkobook} and is
called the ``Replica Symmetry" limit. It is important to clarify
that we are free to choose any form we like for $\lambda_{ab}$, and
that we will later come back to this choice and check whether it is
appropriate or not, in Section~\ref{section_stability}. There is an
important reason why the diagonal elements of the auxiliary field
$\lambda_{ab}$ are chosen to be zero in Eq.~(\ref{RSlambda}). Had
$\lambda_{ab}\,=\,\lambda \mathbbm{1}_{ab}$ been our choice, then
the summation over $a$ and $b$ of $\lambda_{ab}^2$ would have given
$m^2\lambda$. The quadratic dependence on $m$ would have meant that,
after differentiating with respect to $m$ and setting $m=0$, this
term would be equal to zero. Clearly this is not acceptable since
the introduction of the auxiliary field in Eq.~(\ref{Fd_lambda})
requires a non-zero $\lambda_{ab}^2$ quadratic term.

Substituting the replica-symmetric ansatz into the effective
Hamiltonian of Eq.~(\ref{Heff}), we find:
\begin{eqnarray}
\beta H_{\rm eff}= \sum_{a,b} & \bigg\{ & -i \phi\, \delta_{ab}+
N\frac{ \lambda^2}{4\tilde\Gamma}
\left(\mathbbm{1}_{ab}-\delta_{ab}\right)
+\nonumber \\
&&+ \frac{1}{2}\sum_q G_{ab}^{-1}(\bm{n}_a \cdot \bm{n}_b)\bigg\},
\label{h2}
\end{eqnarray}
where the propagator of the $a-b$ replica coupling is given by:
\begin{eqnarray}
G_{ab}^{-1}\,=\,\left( \tilde K q^2 +
2i\phi+2\lambda\right)\delta_{ab} - 2\lambda \mathbbm{1}_{ab}
\label{G^-1}.
\end{eqnarray}
A consequence of replica symmetry is that $G_{ab}^{-1}$ only
involves matrices $\delta_{ab}$ and $\mathbbm{1}_{ab}$, and as a
result its logarithm is easily obtained:
\begin{eqnarray}
\log G_{ab}^{-1}&=&\log \left( \tilde K q^2 +
2i\phi+2\lambda\right)\delta_{ab}+ \nonumber \\
&+&\frac{1}{m} \log \left(1-\frac{2\lambda\,m} { \tilde K q^2 +
2i\phi+2\lambda}\right)\mathbbm{1}_{ab} \label{lnG}
\end{eqnarray}
The path integral over configurations $\bm{n}_a$ with the
statistical weight determined by the effective Hamiltonian
(\ref{h2}) is Gaussian and gives $({\rm Det}
G^{-1})^{-1/2}=\exp(-\frac{1}{2} \,{\rm tr}\log G^{-1})$ for each of
the three vector components of $\bm{n}$. As a result the disorder
free energy is given by:
\begin{eqnarray}
\beta F_d&=&-\frac{1}{2}\frac{\partial}{\partial m} \int d\lambda
\int
d\phi \,\exp\bigg\{ - \frac{N \lambda^2}{4\tilde\Gamma}(m^2-m)+\nonumber\\
&& \left. + i\phi m\,+ \frac{3}{2} \sum_q {\rm tr} \log
G_{ab}^{-1}\bigg\} \right|_{m \rightarrow 0}\label{Fd1}
\end{eqnarray}
from the three (identical) path integrals for the components of
$\bm{n}$.

\subsection{Disorder Free Energy}

The aim of this section is to determine the particular values of
the auxiliary fields ($\lambda_{ab}^*$ and $\phi^*$) that make the
disorder energy of Eq.~(\ref{Fd1}) a minimum. To treat the problem
properly, one would have to evaluate the integrals over
$\lambda_{ab}$ and $\phi$, which is analytically challenging. The
standard way to bypass this difficulty, is to employ the
saddle-point approximation  based on the simple observation that
the exponentially most significant contribution in Eq.~(\ref{Fd1})
will occur when the exponent is a maximum. Therefore
\begin{eqnarray}
\beta F_{d}(K,\,\Gamma) &\approx&
-\frac{1}{2}\frac{\partial}{\partial\,m} \exp
\bigg\{\min_{\lambda,i\phi}\bigg[ -\frac{3}{2}\sum_q tr\log
{G_{ab}^{-1}} \nonumber \\
&&+im\phi - \frac{N \lambda^2}{4\tilde\Gamma}(m^2-m)
\bigg]\bigg\}\bigg|_{m \rightarrow 0} , \label{Fd2}
\end{eqnarray}
where $\min_{\lambda,i\phi}(..)$ represents the minimum of a
function with respect to variations in $\lambda$ and $i\phi$.
After substituting the trace of the logarithm from the
Eq.~(\ref{lnG}), we differentiate with respect to $m$ and then set
$m$ equal to zero. The disorder energy then takes the form:
\begin{eqnarray}
&&\beta F_{d} \approx   \min_{\lambda,i\phi}\bigg\{-i\phi - \frac{N
\lambda^2}{4\tilde\Gamma}+   \label{Fd3} \\
 &+&\frac{3}{2}\sum_q \bigg[\log \left( \tilde K q^2 +
2i\phi+2\lambda\right)-\frac{2\lambda} { \tilde K q^2 +
2i\phi+2\lambda} \bigg] \bigg\} ,\nonumber
\end{eqnarray}
and we are left with the task of finding the stationary point $(i
\phi^*,\lambda^*)$. From now on we move to the continuum limit of
space, where the discrete sum is replaced by $\sum_q \rightarrow
V\int\frac{4\pi q^2}{(2\pi)^3}dq$ and the coherence length $a$ is
taken to zero limit.

\subsubsection{Optimal $i\phi$ in the absence of disorder}

The auxiliary field $\phi$ ensures that the nematic director is a
unit vector. This constraint should of course be enforced whether
the disorder is present or not. In fact, there is no physical
reason why the inclusion of disorder should significantly alter
this constraint. Hence, as a first approximation, we look for the
optimum $i\phi^*$ when there is no disorder in the system. Setting
both $\Gamma$ and $\lambda$ equal zero in the Eq.~(\ref{Fd3}) and
differentiating with respect to $i\phi$ we obtain:
\begin{eqnarray}
&&\frac{\partial F_d}{\partial i\phi}=0 \ \ \Rightarrow  \\
&&1=\frac{3V}{2\pi^2}\bigg[\frac{q_{\rm max}}{\tilde K}-
\frac{\sqrt{2i\phi}}{\tilde K^{3/2}}\tan^{-1}\left(q_{\rm
max}\sqrt{\frac{\tilde K}{2i\phi}}\right)\bigg] . \nonumber
\end{eqnarray}
In the continuous limit of $q_{\rm max}\to\infty$ and the
arctangent is equal to $\pi/2$. This equation can be re-written as
 \begin{eqnarray}
\sqrt{2i\phi}=\frac{4\pi \tilde K ^{3/2}}{3V}\left(-1+ \frac{3q_{\rm
max}V}{2\pi^2 \tilde K}\right) . \label{phi1}
\end{eqnarray}
Clearly the $-1$ term is negligible compared to $(q_{max}V/\tilde
K)$, and can be therefore neglected. Another factor that supports
this omission is that we want to examine what happens close to the
phase transition, where $Q\to 0$. In this limit, the elastic
constant is known to vanish ($\propto Q^2$) and the term with
$\tilde K^{-1}$ dominates in the bracket.

Assuming that the inclusion of disorder has only a minor effect on
$\phi$, the value we will use from now on is:
\begin{eqnarray}
i\phi^*=\frac{2\tilde K q_{\rm max}^2}{\pi^2} \label{phi},
\end{eqnarray}
which tends to infinity in the continuum limit of space. It is
interesting to note that very close to the transition the constraint
relaxes since $\phi \to 0$. This is not surprising since the meaning
of the director itself becomes ill-defined as we approach the
transition point.

The full calculation to obtain $\phi^*(\lambda)$ in a system with
disorder is possible. However, using a disorder-dependent
$\phi^*(\lambda)$ in Eq.~(\ref{Fd3})  makes it analytically
impossible to solve $\partial F_d/\partial\lambda\,=\,0$ and
determine the optimum $\lambda^*$. Even a perturbative approach :
$\phi^*(\lambda)\,=\,\phi^*_{\lambda=0}+\,\it{``small\,
correction"}$ does not help matters. To overcome this difficulty
the $(\phi^*,\lambda^*)$ saddle point should be found numerically,
searching for a global energy minimum in the $Q-\gamma$ space. The
important drawback of the latter is that we will then be unable to
determine the analytical form of $\lambda^*(Q)$ and therefore the
final $F_d(Q)$ correction to the Landau free energy of the phase
transition.

\subsubsection{Optimal $\lambda$ for weak disorder}

We proceed to determine the value of the auxiliary field
$\lambda^*$ that minimizes the energy in the replica symmetric
approximation. Differentiating the right-hand side of
Eq.~(\ref{Fd3}) and demanding it to  be zero we find:
\begin{eqnarray}
\frac{\lambda^* N}{2\tilde \Gamma}=\frac{3V}{\pi^2}\int
\frac{q^2}{\left(\tilde K q^2+ 2i\phi^* + 2\lambda^*\right)^2}\,dq
\, . \nonumber
\end{eqnarray}
Integration over momentum space gives the stationary condition on
the auxiliary field:
\begin{eqnarray}
&&\frac{\lambda^* N}{2\tilde \Gamma}=\frac{3V}{\pi^2}
\bigg[-\frac{q_{\rm max}}{2\tilde K (\tilde K q_{\rm max} ^2+
2i\phi^* +2\lambda^*)} \label{lambda1} \\
&&+ \frac{1}{(2\tilde K)^{3/2}\sqrt{i\phi^*+\lambda^*}}
\tan^{-1}\left(q_{\rm max}\sqrt{\frac{K}{2i\phi^*+2\lambda^*}}
\right)\bigg]
 \nonumber
\end{eqnarray}
In the continuous limit of $q_{\rm max}=2\pi/a \to \infty$ the
first term vanishes and the arctangent is equal to $\pi/2$.
Equation~(\ref{lambda1}) has only one real solution.
Unfortunately, its full expression is too long and cumbersome to
appear here explicitly; instead we demonstrate its behavior in two
limits. Expanded in powers of $\tilde \Gamma \ll 1$ (weak
disorder) it takes the form
\begin{eqnarray}
\lambda^*(\Gamma)=\frac{3V }{N(2\tilde K)^{3/2}\pi\sqrt{i\phi^*}}\,
\tilde\Gamma +\mathcal{O}(\Gamma^2)\label{lambdaG}.
\end{eqnarray}
It is a reassuring property that $\lambda^*$ vanishes as
$\Gamma\to 0$.

Critical to our work is the behavior of $\lambda^*$ as the order
parameter $Q$ tends to zero. Both $\Gamma$ (from its definition)
and $K$ (for small $Q$) are quadratic functions of $Q$. Therefore
the leading term in the series expansion of small $Q$ is
\begin{eqnarray}
\lambda^*(Q)=\frac{(3V)^{2/3}}{2\pi^{2/3}N^{2/3}}\,\frac{\tilde
\Gamma^{2/3}}{\tilde K}+\mathcal{O}(Q^{4/3}) . \label{lambdaQ}
\end{eqnarray}
The scaling $\lambda^*\propto Q^{-2/3}$ is thus obtained, showing
that $\lambda^*$ diverges as the transition is approached, that
is, even a weak disorder becomes relevant near the $Q \rightarrow
0$ point.

\subsubsection{Final disorder free energy}

To find the final disorder energy the values of fields $\phi^*$
and $\lambda^*$ are put back in the Eq.~(\ref{Fd3}). Performing
the $q$-integrations in the continuum limit we obtain:
\begin{eqnarray}
F_d&=&-i\phi^* -\frac{N \lambda^*}{4\tilde \Gamma}+\frac{V}{12
\pi^2}\bigg[-\frac{6 q_{\rm max}(\lambda^*-2i\phi^*)}{\tilde K}
\nonumber\\
&&+ \frac{3\sqrt{2} \pi
\left(\lambda^2-i\phi^*\lambda^*+2{\phi^*}^2\right)}{\tilde K
\sqrt{i\phi^*+ \lambda^* }}\bigg] \label{Fd}
\end{eqnarray}
The energy we are interested in arises from disorder and we can
safely ignore terms that are still present when $\lambda^*=
\Gamma=0$. At diminishing order parameter, the leading term of
Eq.~(\ref{Fd}) takes the form:
\begin{eqnarray}
F_d=\frac{V(\lambda^*)^{3/2}}{2\sqrt{2}\pi\tilde K^{3/2}} \approx
\frac{3 V^2\tilde \Gamma}{8\pi^2 \tilde K^3 N}\propto
Q^{-4}\label{FdQ}\, ,
\end{eqnarray}
which clearly diverges as $Q\to 0$. This divergence of the
disorder free energy implies that the isotropic phase can never be
reached and, as we shall see in greater detail in
Section~\ref{section_analysis}. This in turn leads to the rounding
of the nematic-isotropic phase transition. Before all this is
discussed, we examine if the replica-symmetric form of
$\lambda_{ab}$ in Eq.~(\ref{RSlambda}) was an appropriate choice,
which is far from obvious.

\section{Stability of Replica Symmetry}
\label{section_stability}

\subsection{The Hessian}

A necessary condition for the replica-symmetric solution to be
applicable is that the disorder energy is stable for infinitesimal
variations of that solution. The auxiliary field $\lambda^*$ gives
an energy extremum; whether this extremum is a maximum or a
minimum is determined by a stability analysis following the work
of de Almeida and Thouless in spin glasses \cite{almeida78}. We
start by allowing the matrix of the auxiliary field $\lambda_{ab}$
to deviate from its replica-symmetric form:
\begin{eqnarray}
\lambda_{ab}=\lambda^* \left(\mathbbm{1}_{ab}-\delta_{ab}\right)
+\epsilon_{ab} \ , \label{lambda stability}
\end{eqnarray}
with $\epsilon_{ab}$ the arbitrary infinitesimal deviation from
replica symmetry. The disorder free energy will be expanded to
second order in $\bm{\epsilon}$:
\begin{eqnarray}
F_d(\lambda_{ab})=F_{RS}+\frac{1}{2} \sum_{abcd} H_{ad,
bc}\epsilon_{ab}\epsilon_{cd}\label{stability1}
\end{eqnarray}
where $F_{RS}$ represents the energy of the replica-symmetric case
($\epsilon_{ab}=0$), given by the Eq.~(\ref{Fd}). The second-order
term in $\epsilon_{ab}$ involves the fourth rank tensor of
coefficients $H_{adbc}$, which is called the Hessian. It plays an
important role in this analysis, because the replica-symmetric
solution is only stable as long as $H_{adbc}$ is positive
definite, or equivalently only as long as its eigenvalues are
non-negative. Of course there is no term linear in $\epsilon_{ab}$
because its coefficient is given by the right hand side of
Eq.~(\ref{lambda1}) and is therefore equal to zero.

We proceed to find the Hessian of this model. Expanding the first
term of the disorder energy in powers of $\epsilon_{ab}$ gives:
\begin{eqnarray}
\frac{N}{4\tilde \Gamma}\lambda_{ab}^2=
\frac{N}{4\tilde\Gamma}\left( {\lambda^{rs}_{ab}}^2
+2\lambda^{rs}_{ab}\epsilon_{ab}+\epsilon_{ab}\epsilon_{ab} \right)\
, \label{stab lambda^2 expanded}
\end{eqnarray}
where $\lambda^{rs}_{ab}=\lambda^{*}
\left(\mathbbm{1}_{ab}-\delta_{ab}\right)$. Hence the contribution
to the Hessian from this term is $ (N/2\tilde{\Gamma}) \,
\delta_{ad}\,\delta_{bc}$.

The other contribution comes from the trace of the logarithm. The
propagator of Eq.~(\ref{G^-1}) can be written as
$G^{-1}_{ab}=\left(\tilde{K}q^2+2i\phi\right)\delta_{ab}-2\lambda_{ab}$,
therefore allowing $\lambda_{ab}$ to vary gives the following form
to the propagator
\begin{eqnarray}
G^{-1}_{ab}= {G^{rs}_{ab}}^{-1} -2\epsilon_{ab} \label{Gabstab}\, ,
\end{eqnarray}
where the replica-symmetric ${G^{rs}_{ab}}^{-1}$ is given in
Eq.~(\ref{G^-1}). Therefore,
\begin{eqnarray}
{\rm tr} \, \log G_{ab}^{-1}= {\rm tr} \,
\log\left[{G^{rs}_{ac}}^{-1}\left(\delta_{cb}
-\,2G_{cd}^{rs}\,\epsilon_{db}\right)\right] \
 \label{stabremark1}
\end{eqnarray}
where summation over the dummy indexes $c,d$ is implicit and
\begin{eqnarray}
G^{rs}_{ab}=\frac{1}{\tilde K q^2 +
2i\phi^*+2\lambda^*}\,\delta_{ab} + \frac{ 2\lambda^*}{(\tilde K q^2
+ 2i\phi^*+2\lambda^*)^2} \mathbbm{1}_{ab}  \nonumber
\end{eqnarray}
is the inverse of Eq.~(\ref{G^-1}). In the case of two commuting
matrices: $\log(\bm{A\cdot B})=\log(\bm{A})+ \log(\bm{B})$. It
turns out that the eigenvectors of the Hessian ($\epsilon_{ab}$)
indeed commute with matrix ${G}_{ab}$. This is largely because the
latter is a combination of two very simple matrices
$\mathbbm{1}_{ab}$ and $\delta_{ab}$. Breaking up the product
under the logarithm:
\begin{eqnarray}
\log\bm{G}^{-1}=\log{\bm{G}^{rs}}^{-1}\,+ \log(\bm{\delta}\,
-\,2\,\bm{G}^{rs}\cdot\bm{\epsilon}) \, \nonumber
\end{eqnarray}
and expanding the second term in a Taylor series, we obtain:
\begin{eqnarray}
{\rm tr} \log G^{-1}_{ab}&=& {\rm tr} \log {G_{ab}^{rs}}^{-1}-
2\sum_{ab}G_{ba}^{rs}\epsilon_{ab}\nonumber \\
&&-2\sum_{abcd} G_{da}^{rs}\epsilon_{ab}G_{bc}^{rs}\epsilon_{cd}\ .
 \label{stabtrln}
\end{eqnarray}
There are three identical traces to be considered, one for each
component of the nematic director. Taking into account the $1/2$
factor in front of the Hessian, this contribution is
$-6\sum_{q}G_{da}^{rs}\,G_{bc}^{rs}$ and the overall Hessian takes
the form:
 \begin{eqnarray}
 H_{ad\,bc}=\left(\frac{N}{2\tilde{\Gamma}}\,\delta_{ad}\delta_{bc}-
6\,G_{da}^{rs}G_{bc}^{rs}\right)\, . \label{hessian}
 \end{eqnarray}
This Hessian has three distinct values of its many matrix
elements, depending on how its many indexes are common to the
pairs $\{a,d\}$ and $\{b,c\}$. These values are $H_{aa,bb}=P$,
$H_{aa,bc}=H_{ad,bb}=Q$ and $H_{ad,bc}=R$.

\subsection{Hessian Eigenvalues}

The eigenvalues of such a 4-rank tensor have been computed by de
Almeida and Thouless in their classical work on replica symmetry
breaking in spin glasses \cite{almeida78}. For $m\to 0$ only two
distinct eigenvalues exist:
\begin{eqnarray}
\Lambda_1=P-4Q+3R \quad {\rm and}\quad \Lambda_3=P-2Q+R \,
\label{eigenvalues}
\end{eqnarray}
and it is straightforward to obtain them by inserting the
appropriate forms of $\bm{G}^{rs}$ into the expression of the
Hessian of Eq.~(\ref{hessian}). The first eigenvalue
\begin{eqnarray}
\Lambda_1&=&\frac{N}{2\tilde{\Gamma}}- 6V \int \frac{4\pi
q^2}{(2\pi)^3} \left[\frac{1}{(\tilde{K}q^2+2i
\phi+2\lambda)^2}\right.\nonumber \\
&& \qquad \qquad -
\left.\frac{4\lambda}{(\tilde{K}q^2+2i\phi+2\lambda)^3}\right]dq
\nonumber \\
 &=&\frac{N}{2\tilde{\Gamma}}-\frac{3(2i\phi+\lambda)V}
{8\sqrt{2}\pi\tilde{K}^{3/2}(i\phi+\lambda)^{3/2}}
\label{eigenvalue1}
\end{eqnarray}
is degenerate and corresponds to two eigenvectors
$\bm{\epsilon}_1$ and $\bm{\epsilon}_2$. The first,
$\bm{\epsilon}_1$, is symmetric under interchange of indexes
($\epsilon_{ab}=\alpha$ for all $a,b$) and determines whether the
replica-symmetric fixed point of Eq.~(\ref{lambda1}) is stable or
not. In other words, if $\Lambda_1<0$ the replica-symmetric
solution corresponds to an energy maximum and has no physical
relevance.

The remaining two eigenvectors check the general stability of the
replica-symmetric scheme. Contrary to $\bm{\epsilon}_1$, the
second eigenvector $\bm{\epsilon}_2$, corresponding to the
degenerate $\Lambda_1$, is symmetric under interchange of all but
one index ($\epsilon_{ab}=\beta$ for $a$ or $b$ = $c$ and
$\epsilon_{ab}=\gamma$ otherwise). This eigenvector is not
symmetric and as a result a negative $\Lambda_1$ also means that
replica symmetry must be broken to determine the correct
$\lambda_{ab}$. The second eigenvalue is non-degenerate and is
given by:
\begin{eqnarray}
\Lambda_3&=&\frac{N}{2\tilde{\Gamma}}- 6V \int \frac{4\pi
q^2}{(2\pi)^3} \frac{1}{(\tilde{K}q^2+2i \phi+2\lambda)^2}dq=
\nonumber \\
 &=&\frac{N}{2\tilde{\Gamma}}-\frac{3V}
{4\sqrt{2}\pi\tilde{K}^{3/2}(i\phi+\lambda)^{1/2}}\, .
\label{eigenvalue2}
\end{eqnarray}
and its corresponding eigenvector, called $\bm{\epsilon}_3$, also
does not have a symmetric form ($\epsilon_{ab}=\delta$ for $a=c$
and $b=d$, $\epsilon_{ab}=\zeta$ for $a=c$ or $a=d$ and $b\neq
c,d$, $\epsilon_{ab}=\eta$ otherwise). Similarly to the previous
case, if $\Lambda_3<0$ then the replica symmetric solution breaks
down and different forms of $\lambda_{ab}$ must be sought.

Accordingly, a sufficient condition for the replica-symmetric
solution to be stable is that all the eigenvalues of the Hessian
remain positive. Both $\Lambda_1$ and $\Lambda_3$ show a similar
behavior as functions of disorder strength $\gamma$, which was
defined in the text above Eq.~(\ref{rfenergy}), and the order
parameter $Q$. Substituting the optimal $\phi^*$ and $\lambda^*$,
we find that when $Q$ does not tend to zero, both eigenvalues
remain positive, especially for weak disorder, $\tilde{\Gamma}\ll
1$. As the continuous phase transition is approached and the order
parameter diminishes, the three parameters appearing in
Eq.~(\ref{eigenvalue2}) become simple functions of $Q$:
$\Gamma\propto Q^2$, $K\propto Q^2$ and $\lambda^* \propto
Q^{-2/3}$. Hence, the eigenvalues scale as $Q^{-2}-Q^{-8/9}$ and,
therefore, become negative at some point $Q \ll 1$. This means
that for $Q \to 0$ the replica-symmetric solution of the auxiliary
field Eq.~(\ref{lambdaQ}) and the resulting expression for the
disorder energy, Eq.~(\ref{FdQ}), correspond to an energy maximum
and should not be used.

To find the exact point at which replica symmetry becomes unstable
we note that it is the second (non-degenerate) eigenvalue,
$\Lambda_3$ that becomes negative first as $Q\to 0$. Substituting
the parameters $\Gamma$, $K$ and $\lambda^*$ we find this crossover
value for stability:
\begin{eqnarray}
Q_{\rm stab}=\frac{3g}{4\sqrt{2}\pi(\beta \kappa a )^{3/2}} \, ,
\label{Qstab}
\end{eqnarray}
where the relation $V/N\,=a^3$ has been used, with $a$ the
short-length cutoff. Bearing in mind that $g\propto a^3$ and that
$\kappa\propto 1/a$ we find $Q_{stab}\propto a^3$. To be
consistent with previous calculations the continuum limit of space
is employed and $a\to 0$. Therefore the threshold value below
which replica symmetry breaks tends to zero and the
Eq.~(\ref{lambdaQ}) is always valid in the continuum limit.
Another way to see this is by examining the explicit dependance of
$\Lambda_3$ on the cut-off length-scale $a$. The first term
$(N/\tilde{\Gamma})$ scales as $a^{-6}$ whereas the second term
scales as $-a^{-5}$. Therefore when $a$ is taken to zero the first
term dominates and the eigenvalue remains positive. To complete
the stability analysis, we note that the other eigenvalue also
becomes negative for smaller values of $Q$ when $Q<Q_{stab}/2$,
but this is irrelevant since the instability of $\Lambda_3$ occurs
first.

\section{Phase Transition Analysis}
\label{section_analysis}


The aim of this paper is to discuss how quenched orientational
disorder affects the phase transition of nematic systems. The
total free energy density is a combination of the Landau-De Gennes
expansion of the order parameter plus the disorder part ($F_d/V$)
obtained in Eq.~(\ref{Fd}). To get the total energy as an
expansion of the order parameter only the leading order
contribution of $F_d$ in Eq.~(\ref{FdQ}) is considered:
\begin{eqnarray}
F= \frac{A_0}{2} (T-T^*)Q^2-\frac{B}{3}\,Q^3+\frac{C}{4}\,Q^4
+\frac{D}{4}\, Q^{-4}\,,
 \label{energy}
\end{eqnarray}
where
\begin{eqnarray}
 D=\frac{3g}{2\pi^2(\kappa\beta a^2 )^3}k_BT
 \label{D}
 \end{eqnarray}
and $A$, $B$, $C$ and $T^*$ are the usual Landau-De Gennes
parameters. One may note the explicit dependence of Eq.~(\ref{D})
on the short-length cutoff $a$ and be concerned about the
sustainability of the continuum limit $a\to 0$. In fact, because
$g \propto a^3$ and $\beta \kappa \simeq 1/a$, the powers of $a$
cancel and the free energy does not depend on this cutoff
parameter, which is a reassuring test of consistency of our
theory.

\begin{figure}[tbp]
\begin{center}
\resizebox{0.4\textwidth}{!}{\includegraphics{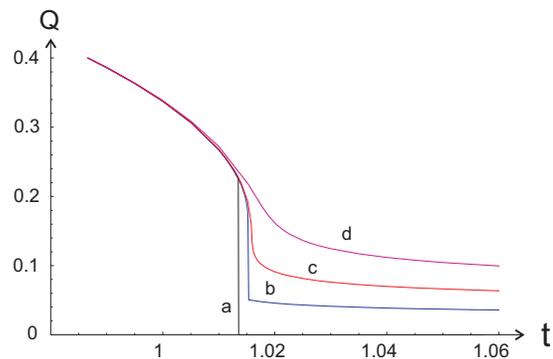}}
\end{center}
\caption{The equilibrium order parameter $Q$ as a function of
reduced temperature $t= {T}/{T^*}$ for a range of different
disorder strength $g$. (a) is the first order transition for a
system with no disorder (b) is a subcritical system (c) is a
critical system (d)  is a supercritical system. As the disorder
strength increases the discontinuous jump decreases and eventually
disappears.} \label{plotQofT}
\end{figure}

The easiest way to illustrate the consequence of the free energy
renormalization is to plot the equilibrium values of the order
parameter $Q$. The values chosen for the Landau phenomenological
constants are: $A_0\simeq 5.0 \times 10^3\,\hbox{Jm}^{-3}K^{-1}\,,
\ B\simeq 3.3 \times 10^5\,\hbox{Jm}^{-3}\ {\rm and}\ C\simeq1.0
\times 10^6 \hbox{Jm}^{-3}$. A detailed description on how they
are obtained by analyzing experimental data of Ref.~\cite{disch94}
is given in the Appendix. We consider these values to be
indicative only since different nematic materials will certainly
have large variations in these values. Plotting the equilibrium
order parameter against reduced temperature, Fig.~\ref{plotQofT},
we can compare $Q(T)$ for the disorder strength increasing from
$g=0$ in (a) up to $g=10^{-4}$ in (d). As mentioned in
Section~\ref{section_3} a typical polydomain elastomer should have
$g\simeq 4\times 10^{-3}$ and therefore the model predicts a
supercritical behavior in agreement with many experiments. Note
that in contrast to this result, the quenched orientational
disorder was shown not to alter the continuous nematic phase
transition in thin films, whose director is confined in the
$XY$-plane \cite{petridis06}.

The inclusion of disorder has a profound effect on the phase
behavior of 3D systems whose pure versions undergo a first order
phase transition. The discontinuous jump of the order parameter at
the nematic transition becomes smaller as the strength of the
disorder $g$ increases and eventually disappears altogether above
a critical value making the phase transition continuous. The
change in behavior is explained by the simple fact that the energy
terms arising from disorder scale as negative powers of $Q$. As a
direct result, the energy of the system increases as the
transition approaches and the zero order parameter phase is never
reached. To see exactly how this happens consider plots of $F(Q)$
of a subcritical and a supercritical system.

\begin{figure}[tbp]
\begin{center}
\resizebox{0.35\textwidth}{!}{\includegraphics{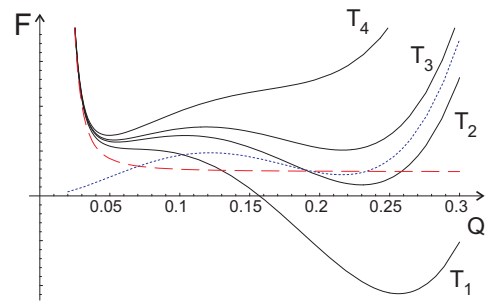}}
\end{center}
\caption{Free energy against order parameter plots for a
subcritical
 system for a range of temperatures, $T_1\,<\,T_2\,<\,T_3\,<\,T_4$.
The doted and dashed lines show respectively the Landau and
disorder energies for $T=T_3$. They illustrate that the high-Q
minimum is a product of the Landau energy exclusively, whereas the
low-Q minimum arises from competition of the $A\,Q^2$ Landau term
and the disorder energy.
 } \label{fig jump}
\end{figure}

For a subcritical system, where disorder is weak, the jump in
$Q(T)$ is still present, albeit smaller than in the original
transition of the pure system. Because parameter $g$ is small, the
disorder part of the energy becomes significant only for small
$Q$, where it diverges. The appearance of the jump has the same
origins as in the classical Landau-De Gennes theory. Fig.~\ref{fig
jump} shows plots of the energy density against order parameter
for four different temperatures around $T_{NI}$. At the lowest
temperature ($T_1$) the single minimum determines the equilibrium
value of $Q$. As the temperature increases the high-$Q$ minimum
moves to a slightly smaller value of $Q$ and -more importantly-
another ``low-Q" minimum appears as a result of disorder. At $T_2$
the high-$Q$ is still the global minimum, but at the critical
temperature ($T_3$) the two minima have the same energy value.
This means that two distinct phases, one with $Q\simeq 0.23$ and
the other $Q\simeq 0.05$, coexist. Once this temperature is passed
($T_4$) the low-$Q$ minimum determines the system's order
parameter. The crucial difference between these plots and the
classical Landau-De Gennes theory is that the low-$Q$ minimum in
the latter is always placed at $Q=0$. Since the disorder energy
diverges at zero $Q$, this minimum is pushed at positive values of
$Q$ in systems with quenched disorder.

The dotted and dashed lines show the Landau and the disorder
energy for the same temperature $T_3$, respectively. Around the
high-$Q$ minimum the dotted line has the same shape as the actual
energy; apart from a constant shift to lower energy they are
exactly equal. Therefore this minimum is a result of the
competition of the $-B\,Q^3$ and $C\,Q^4$ energy terms of the
Landau expansion, which dominate at large $Q$. The position of the
low-$Q$ minimum is influenced by disorder. This minimum is a
balance of the divergent $F_d$ term and $A\,Q^2$. For temperatures
well above $T_{NI}$ the minimum is located at very small order
parameter and the only relevant terms of Eq.~(\ref{energy}) are
$DQ^{-4}$  and $A\,Q^2$.

\begin{figure}[tbp]
\begin{center}
\resizebox{0.35\textwidth}{!}{\includegraphics{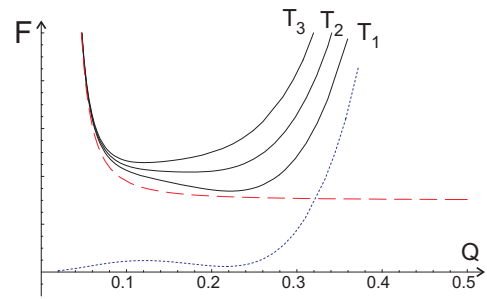}}
\end{center}
\caption{Energy against order parameter plots of a supercritical
system for a range of temperatures, $T_1\,<\,T_2\,<\,T_3$, all
larger than $T_{NI}$. The dotted  and dashed lines show the Landau
and disorder energies for $T_1$, respectively. Since the disorder
energy diverges the $A\,Q^2$ Landau term is no longer significant.
 } \label{fig nojump}
\end{figure}

In a supercritical system disorder is stronger (large $g$) and
therefore its effect on the energy is more prominent. As a result
the effect of $F_d$ is relevant for all the values of $Q$, not
just in the small order parameter region as in the previous case.
 Fig.~\ref{fig nojump} shows the relevant energy plots. Crucially there is only one
minimum at any given temperature. Its position shifts to smaller
order parameter as temperature increases, but the phase transition
is continuous. In comparison with Fig.~\ref{fig jump}, we can say
that the low-$Q$ minimum has ``broadened" and ``absorbed" the
high-$Q$ minimum.

Another difference with the small $g$ case is that, because $F_d$
has larger magnitude, the low order $A\,Q^2$ term of the Landau
expansion does not influence the position of the minimum. The dotted
lines in Fig.~\ref{fig nojump} are drawn for the same temperature as
in the previous case of Fig.~\ref{fig jump}, but now this
temperature is labeled as $T_1$. In this supercritical system the
energy minimum occurs at approximately the same value of order
parameter as the high-$Q$ minimum of the Landau DeGennes expansion
(thin-doted line). In a pure system this would be a metastable state
because there would exist a global minimum at $Q=0$.

The natural question to ask next is what is the critical point at
which the jump in $Q(T)$ disappears completely. At this point the
free energy must have the merging of all its minima and maxima.
Solving for the first, second and third derivatives being zero
provides three equations for the unknown critical parameters
$g_c$, $T_c$ and $Q_c$ and the critical point is then given by:
\begin{eqnarray}
(g_c,Q_c)=\left(\frac{2\pi^2 7^{11}B^{12}(\kappa\beta a)^3}{9
16^{12}C^{11}}\,,\,\frac{7B}{16C}\right)\,.
 \label{critical}
\end{eqnarray}
It is worth comparing this with the classical case of an applied
field, which adds a $-f\,Q$ term in the Landau expansion, where
the critical $Q$ is slightly smaller and equal to $B/3C$.

In the analysis of the replica symmetry stability in
Section~\ref{section_stability} we found that the above
description of nematic systems breaks down for small order
parameter when $Q<Q_{\rm stab}$, see Eq.~(\ref{Qstab}). The
threshold value of $Q_{\rm stab}$ is proportional to the cube of
the cut-off length $a$. In order to be consistent with the line
adopted in previous calculations, the continuum limit of space was
employed and $a$ was taken to zero. Hence $Q_{\rm stab}$ was also
taken to be zero. However, in reality the length scale $a$ is in
fact the nematic coherence length, because below this size we
cannot write Frank elasticity and there is no meaning to order
parameter $Q$, or director $\bm{n}$. Keeping $a$ non-zero, and
assigning it the value of $5$nm which is usually associated with
liquid crystals, makes $Q_{\rm stab}$ also finite. Nevertheless
substituting the value of $g \simeq 3\times 10^{-3}$ and
$\kappa\simeq k_BT/a$ gives $Q_{\rm stab}\simeq 2\times 10^{-4}$.
This is an extremely small value for the order parameter and, as
we see from Fig~\ref{plotQofT}, it would only be acquired at
temperatures well above the experimental range. Hence, even if
$a\neq 0$, the window of replica symmetry stability is wide enough
to describe realistic nematic systems.

\section{Summary}

This paper examines how the inclusion of randomly quenched
orientational disorder leads to the rounding of the
nematic-isotropic phase transition in three dimensions. The coupling
between impurities and the local order parameter pins some mesogenic
molecules and does not allow the sample to have a uniform director
field $\bm{n}(r)$. After quenched disorder has been averaged over
using the replica method, a replica-symmetric auxiliary field is
used to obtain the free energy arising from disorder. The disorder
energy adds a $\propto Q^{-4}$ term to the Landau-De Gennes
expansion which diverges for diminishing order parameter $Q$. As a
result the isotropic phase is never reached and, for sufficiently
strong disorder, the phase transition becomes continuous. This is in
accordance with many experiments on nematic elastomers that also
show a smooth transition rather than a discontinuous as predicted by
the classical Landau-De Gennes theory. An earlier study on XY
nematics, whose director is confined to a plane, showed that the
quenched disorder does not affect their continuous phase behavior
\cite{petridis06}.

A stability analysis shows that the replica-symmetric solution we
have employed does fail at small values of the order parameter, at
$Q<Q_{stab}$. However, for realistic values of physical parameters
we estimate the order of $Q_{stab}\approx 10^{-4}$ in nematic
elastomers. A supercritical system acquires such low values of $Q$
at temperatures well above the elastomers melting point and
therefore the window of replica symmetry stability adequately
describes the 3D nematic system.

\acknowledgments

We would like to thank Isaac P\'erez Castillo, David Sherrington,
Paolo Biscari and Raphael Blumenfeld for their feedback. This work
has been supported by the Leventis foundation, the Cambridge
European Trust and the EPSRC TCM/C3 Portfolio grants.

\appendix*
\section{}

In order to determine the three phenomenological parameters $A_0$,
$B$ and $C$ of the Landau-De Gennes theory, defined in
Eq.~(\ref{energy}), we need three independent measurements. These
are provided by a NMR experiment measuring the order parameter as
a function of temperature of a polymer melt and its corresponding
crosslinked network (a polydomain nematic elastomer)
\cite{disch94}. As expected from the Landau-De Gennes theory, the
melt shows a first order transition with the discontinuous jump in
order parameter being approximately $\Delta Q\simeq 0.22$, which
is smaller than $0.4$, the value usually associated with ordinary
liquid crystals \cite{shen73}. The theoretical prediction gives
this jump to be equal to $\Delta Q=2B/3C$ \cite{degennes95} and
substituting the experimental measurement we find $B\simeq 0.33C$.
The second measurement is the width of the temperature hysteresis
which is $\Delta T\simeq 5K$ in the polymer melt. Theory predicts
$\Delta T=2B^2/9A_0C$. Combining this with $B\simeq C/3$ we find $
B= 67\,A_0\ 1K$ and $C=200\,A_0 \ 1K$. To get a value of $A_0$ a
third measurement is required.

A striking difference between the two $Q(T)$ plots, for a nematic
polymer melt and its crosslinked elastomer version, is that
crosslinking has reduced the overall order at temperatures below
$T_{NI}$. For elastomers crosslinked in the isotropic phase, the
energy addition arising from nematic rubber elasticity adds a
fourth order term in the Landau expansion \cite{warner88}:
\begin{eqnarray}
\frac{3}{4}\,\mu\,\alpha^4\,Q^4 \nonumber
\end{eqnarray}
where $\mu$ is the rubber modulus and $\alpha$ accounts for the
microscopic details of an elastomer. For a freely joined polymer
$\alpha=3$ \cite{emtbook}, but a side-chained polymers have $\alpha$
ranging between $-0.5$ and $0$ \cite{mitchell92}. Let us take an
intermediate case where $\alpha=1$. When this term is added to the
Landau-De Gennes expansion, the fourth-order coefficient (which is
$C/4$ for the polymer melt) now becomes larger
$\frac{C}{4}(1+3\mu\alpha^4/C)$. Hence the transition temperature,
given by $T_{NI}=T^*+2B^2/3A_0C^2$, decreases since the renormalized
$C$ increases. The shift in this transition temperature between the
melt and the corresponding elastomer is:
\begin{eqnarray}
\Delta T_{NI}\,=\,-\frac{2\,B^2\,\mu\,\alpha^3}{3\,A_0\,C^2}\ .
\label{delta TNI}
\end{eqnarray}
and it provides the third relation that allows to determine $A_0$.
An estimate of $\Delta T_{NI}$ is possible in Ref~\cite{disch94}:
$T_{NI}$ is easily identified in the polymer melt, but it is not
clear what it means in the disordered nematic elastomer with a
continuous transition.  It can be loosely defined as the
temperature where $Q=\Delta Q=0.22$. This then makes $\Delta
T_{NI}\simeq 15^o\,C$. Substituting this back to Eq.~(\ref{delta
TNI}) with $\alpha=1$, we obtain $A_0\,\simeq\,\frac{\mu}{200\
K}$. A typical nematic elastomer has elastic modulus of the order
of $10^6$\,Pa. Putting everything together, the phenomenological
constants of a nematic elastomer are:
 \begin{eqnarray} && A_0 \simeq 5.0 \times
10^3\,\hbox{Jm}^{-3}K^{-1}, \ \   B\simeq 3.3 \times
10^5\,\hbox{Jm}^{-3} \nonumber \\
&& \ \ \ {\rm and} \ \  C\simeq 1.0 \times 10^6 \hbox{Jm}^{-3}.
\nonumber
 \end{eqnarray}
These values are crude estimations, only given here to illustrate
the effect of disorder in our model. However, it is comforting that,
although obtained from a different set of experimental measurements,
these values are quite close to the ones reported in literature
\cite{emtbook}.


\end{document}